\documentclass[12pt]{article}
\usepackage{amsfonts}
\usepackage{amsmath, amssymb}
\usepackage{youngtab}
\usepackage{hyperref}
\usepackage[dvips]{color}
\usepackage{psfrag}
\usepackage{feynmp}
\usepackage[pdftex]{graphicx}
\usepackage{braket}
\usepackage{bm}
\usepackage{subfigure}

\usepackage{comment}
\includecomment{pdffig}

\allowdisplaybreaks[1]

\Yboxdim{5pt}

\newcommand{\bel}{\begin{eqnarray}}
\newcommand{\ee}{\end{eqnarray}}

\def\rem#1{}

\makeatletter
    
    \@addtoreset{equation}{section}
  \makeatother

\usepackage{color}

\def\rem#1{}

\renewcommand{\title}[1]{\vbox{\center\LARGE{#1}}\vspace{5mm}}
\renewcommand{\author}[1]{\vbox{\center\large#1}\vspace{5mm}}

\begin{document}
\bibliographystyle{utphys}

\begin{titlepage}
\begin{center}
\vspace{5mm}
\hfill {
}
\\
\vspace{20mm}

\title{
\Large
\bf  
Fermionic extensions of $W$-algebras via  3d $\mathcal{N}=4$    gauge theories with a boundary
}
\vspace{7mm}

{
Yutaka Yoshida
\\
}

\vspace{6mm}

\vspace{3mm}
{\small {\it Department of Current Legal Study, Faculty of Law, Meiji Gakuin University, 1-2-37 
Shirokanedai, Minato-ku, Tokyo 108-8636, Japan}} \\
{\small {\it Institute for Mathematical Informatics, Meiji Gakuin University
1518 Kamikurata-cho, Totsuka-ku, Yokohama 244-8539, Japan}}

\end{center}

\vspace{7mm}
\abstract{
We study properties of vertex (operator) algebras associated with 3d H-twisted $\mathcal{N}=4$ supersymmetric gauge theories with a boundary. The vertex operator algebras (VOAs) are defined by BRST cohomologies of currents with symplectic bosons, complex fermions, and bc-ghosts. We point out that VOAs for 3d $\mathcal{N}=4$ abelian gauge theories are fermionic extensions of VOAs associated with toric hyper-K\"{a}hler varieties.  From this relation, it follows that the VOA associated with the 3d mirror of $N$-flavor $U(1)$ SQED is a fermionic extension of a $W$-algebra $W^{-N+1}(\mathfrak{sl}_N, f_{\text{sub}})$. For $N=3$, we explicitly compute the OPE of elements in the BRST cohomology and find a new algebra that is a fermionic extension of a Bershadsky-Polyakov algebra $W^{-2}(\mathfrak{sl}_3, f_{\text{sub}})$.  We also suggest an expression for the vacuum character of the fermionic extension of $W^{-N+1}(\mathfrak{sl}_N, f_{\text{sub}})$ predicted by 3d $\mathcal{N}=4$ mirror symmetry.
 }
\vfill

\end{titlepage}

\tableofcontents

\section{Introduction}
The vertex  operator algebras (VOAs) are  important mathematical objects obtained by axiomatizing the properties of operator product expansion (OPE) in two-dimensional (2d) conformal field theory. It has applications not only in other areas of mathematics but also in various fields such as string theory and supersymmetric quantum field theory.
  In recent years, there has been growing interest in studying VOAs associated with 
four-dimensional (4d) $\mathcal{N}=2$ superconformal field theories \cite{Beem:2013sza, Beem:2014kka, Beem:2014rza}.

Three-dimensional (3d) $\mathcal{N}=4$ supersymmetric gauge theories have received considerable attention due to their relevance in  
infrared dualities, superstring theory  and other areas of mathematical physics such as AdS/CFT correspondence. 
Costello and Gaiotto \cite{Costello:2018fnz} introduced a family of vertex (operator) algebras  associated with 3d $\mathcal{N}=4$ supersymmetric gauge theories with the H-twist. 
These VOAs, denoted by $V_H(\mathcal{T})$, are defined using the data of a supersymmetric gauge theory $\mathcal{T}$  and  living on
 the 2d boundary of a 3d spacetime. 
The main idea behind the construction of  $V_H(\mathcal{T})$ is to consider the space of observables of the supersymmetric gauge theory on the boundary of the spacetime. This space   is equipped with a natural  OPE that encodes the algebraic relations among the observables. 
In more detail, the VOA $V_H(\mathcal{T})$ is defined as a quotient of symplectic bosons, complex fermions, and bc-ghosts. The symplectic bosons encode the algebra of hypermultiplet scalars. 
The fermions are identified with the fermions in the $\mathcal{N}=(0,2)$  fermi multiplets introduced to   
cancel the gauge anomaly by the anomaly inflow mechanism. The quotient is taken by a BRST cohomology with bc-ghosts associated with the 3d vector multiplet and corresponds to picking up gauge invariant operators.

VOAs associated with 3d $\mathcal{N}=4$ gauge theories offer new insights into the geometric and algebraic structure of these theories, including 3d mirror symmetry that exchanges the Higgs and Coulomb branches between dual theories. 
They would provide a new tool for studying the representation theory of these theories and their physical observables. 

In this paper, we investigate the construction and properties of VOAs associated with 3d $\mathcal{N}=4$  supersymmetric abelian gauge theories with  the H-twist. 
We  explore the relation between the VOAs associated with 3d $\mathcal{N}=4$ abelian gauge theories and VOAs  associated with toric hyper-K\"ahler varieties introduced by Kuwabara \cite{MR4320811}.
We also study the algebraic structure of the VOA associated with the 3d mirror of $U(1)$ $N$-flavor SQED and show that the VOA is a fermionic extension of a $W$-algebra $W^{-N+1}(\mathfrak{sl}_N, f_{\rm sub})$. 

The paper is organized as follows. 
In section \ref{sec:section2}, we review the basics of VOAs associated with 3d $\mathcal{N}=4$ supersymmetric abelian gauge theories and study their relation to the VOA associated with toric hyper-K\"ahler varieties.
In section \ref{sec:section3}, we describe the construction of VOAs associated with the mirror of SQED and show that the VOA has $W$-algebra $W^{-N+1}(\mathfrak{sl}_N, f_{\rm sub})$ as a sub-algebra. For $N=3$,  we  compute the OPE of an expected generator of the BRST cohomology defining the VOA and show that the OPE is closed. 
 In section \ref{sec:section4}, we discuss  applications of supersymmetric indices on a solid torus $S^1 \times D^2$ in understanding  properties of the generator and the vacuum character of the VOA. Finally, in section \ref{sec;section5}, we summarize our results and discuss future directions of research.

\section{VOA associated with 3d $\mathcal{N}=4$ abelian  gauge theories}
\label{sec:section2}
\subsection{Definition}

Costello-Gaiotto  introduced VOAs  associated with  3d $\mathcal{N}=4$ H-twisted supersymmetric gauge theories  on a spacetime $\mathbb{R}^2 \times \mathbb{R}_{\le 0}$.  First we briefly review the construction of VOA for an abelian gauge theory with
a gauge group $G=\prod_{a=1}^L U(1)_a$ \cite{Costello:2018fnz, Costello:2018swh} and $N$ hypermultiplets.
A 3d $\mathcal{N}=4$ hypermultiplet consists of  two 3d $\mathcal{N}=2$ chiral multiplets.  
Let  $(q_i, \tilde{q}_i)$  be the scalar component of two chiral multiplets in the $i$-th hypermultiplet which transform  as  $(q_i, \tilde{q}_i) \mapsto   (e^{{\rm i} Q_{a, i} \alpha }q_i, e^{-{\rm i} Q_{a, i} \alpha } \tilde{q}_i)$ under the $U(1)_a$-gauge transformation,  $a=1,\cdots , L $, and $ i=1,\cdots , N$.

When the spacetime has a boundary, we have to specify  boundary conditions for the fields. 
We impose the $\mathcal{N}=(0,2)$ Neumann boundary conditions for two 3d $\mathcal{N}=2$ chiral multiplets in each hypermultiplet,
and impose  the $\mathcal{N}=(0,2)$ Neumann (resp. Dirichlet) boundary conditions for 3d $\mathcal{N}=2$ vector (resp. adjoint chiral) multiplet in the 3d $\mathcal{N}=4$ $G$ vector multiplet. Then the 3d $\mathcal{N}=(0,4)$ supersymmetry is preserved at the boundary.
Here ``Neumann'' and ``Dirichlet''  are  a terminology  used in  \cite{Yoshida:2014ssa, Dimofte:2017tpi}.
The above boundary condition preserves $ \mathfrak{su}(2)_H$ R-symmetry in the 3d $\mathcal{N}=4$ supersymmetry algebra. In particular, the H-twist, which is a twist by    $\mathfrak{u}(1)_{\mathrm{rot}} \times \mathfrak{u}(1)_{H}  \subset \mathfrak{su}(2)_{\mathrm{rot}} \times \mathfrak{su}(2)_{H}$, makes sense. Here $\mathfrak{su}(2)_{\mathrm{rot}}$(resp. $\mathfrak{u}(1)_{\mathrm{rot}}$) is the Lie algebra of the rotation group acting on  a spacetime  $\mathbb{R}^3$ (resp. $\mathbb{R}^2 \times \mathbb{R}_{\le 0}$).  

In  3d $\mathcal{N}=4$ gauge theories with the H-twist, the gauge and flavor Chern-Simons terms are 
generated by the one-loop fermion effects. When the spacetime has a boundary, the  gauge symmetry is broken by  the boundary term generated by the gauge transformation of the Chern-Simons term. 
In order to cancel the boundary term  by the anomaly inflow, we introduce $\mathcal{N}=(0,2)$ fermi multiplets which coupled to the  $G$ gauge field in three dimensions at the boundary.
Then fermions in the fermi multiplets are identified with 
the complex fermions in the VOA. 
For the abelian gauge group $G$, there is a canonical gauge charge assignment for the fermi multiplets\footnote{When the group $G$ is non-abelian,  a gauge representation  of  $\mathcal{N}=(0,2)$ fermi multiplet satisfying the anomaly inflow highly depend on a choice of $G$ and a gauge  representation of the hypermultiplet.}. To see this, we introduce $\tilde{N}$-flavors  fermi multiplets with gauge charges $\tilde{Q}_{a, i}$, $a=1,\cdots , L $ and $ i=1,\cdots , \tilde{N}$.

The BRST cohomology is defined as follows.
First we associate symplectic bosons $(X_i, Y_i)$
  with the hypermultiplet  scalars $(q_i, \tilde{q}_i)$ having  the following OPE: 
\begin{align}
X_{i}(z) Y_j (0)  \sim \frac{\delta_{i j}}{z}\,.
\label{eq:OPEsb}
\end{align}
A current $J^a_{\rm sb}$ of the symplectic bosons $(X_i, Y_i), i=1,\cdots,N$ is defined by the complex moment map  
$\mu^{a}_{\mathbb{C}}$ for the superpotential of the 3d $\mathcal{N}=4$ abelian gauge theory:
\begin{align}
J^a_{\rm sb} =\mu^{a}_{\mathbb{C}}(X,Y) =\sum_{i=1}^N Q_{a, i} X_{i} Y_i\,. 
\label{eq:currentsb}
\end{align}

 We associate complex fermions $(\psi_i, \chi_i)$ , $i=1, \cdots, \tilde{N}$  with the fermi multiplets and define the OPE by
\begin{align}
\psi_{i}(z) \chi_j (0)  \sim \frac{\delta_{i j}}{z} \,,
\end{align}
and define the current $J^a_{f}$ of the fermions by
\begin{align}
J^a_{\text{f}}:=\sum_{j=1}^{\tilde{N}} \tilde{Q}_{a, i}  \psi_{j}\chi_j\,.  
\end{align}
Note that  the OPE of the fermion currents is given by
\begin{align}
J^a_{\text{f}} (z ) J^b_{\text{f}} (0) \sim \frac{\sum_{i=1}^N \tilde{Q}_{a, i} \tilde{Q}_{b, i}}{z^2}\,.
\label{eq:fcuurent}
\end{align}
In this paper, we used a Mathemtica package \texttt{OPEdefs} \cite{Thielemans:1991uw} to calculate OPE. 
 We introduce $\oplus_{a=1}^L \mathfrak{u}(1)_a$ bc-ghost $({\sf b}_a, {\sf c}_a), a=1,\cdots, L$  associated with the $\prod_{a=1}^L U(1)_a$ vector multiplet, which have the following OPE:
\begin{align}
{\sf b}_{a}(z) {\sf c}_b (0)  \sim \frac{\delta_{a b}}{z}.
\label{eq:bcghost}
\end{align}

The  BRST charge $Q_{\rm BRST}$ and BRST current $J_{\rm BRST}$ are defined by 
\begin{align}
Q_{\mathrm{BRST}}= \oint \frac{dz}{2 \pi {\rm i}}  J_{\rm BRST}(z) \,, \,\,\,
J_{\rm BRST}(z):=\sum_{a=1}^L  {\sf c}_a  \left(  J^a_{\rm sb}(z)+ J^a_{\rm f}(z) \right)\,.
\label{eqBRSTQ}
\end{align}
To define  the BRST cohomology,  the BRST charge has to be nilpotent $Q_{\rm B RST}^2=0$.  
The OPE of the BRST current is given by
\begin{align}
J_{\rm BRST}(z) J_{\rm BRST}(0) \sim  \frac{1}{z}\sum_{a=1}^L \left( \sum_{i=1}^N (Q_{a, i})^2-\sum_{i=1}^{\tilde{N}} (\tilde{Q}_{a, i})^2 \right) \,.
\label{eq:OPEJBJB}
\end{align}
Here  the  OPE coefficient of the BRST current:
$\sum_{i=1}^N (Q_{a, i})^2-\sum_{i=1}^{\tilde{N}} (\tilde{Q}_{a, i})^2$ is same as the coefficient of  the boundary $U(1)_a$-gauge anomaly induced by the 3d fermions in the hypermultiplets and the  boundary fermi multiplets.
When the gauge anomalies  cancel out, the BRST charge becomes nilpotent.
For the abelian gauge theories, there is a canonical choice of the charges and the number of the fermi multiplets:
\begin{align}
\tilde{Q}_{a, i}= {Q}_{a, i}, \quad \tilde{N}=N \,.
\label{eq:anomaly}
\end{align} 
From now on, the  gauge charge of the fermi multiplets are taken as \eqref{eq:anomaly}.
 
Then the vertex (operator) algebra $V_H(\mathcal{T})$ associated with a 3d H-twisted $\mathcal{N}=4$ gauge theory $\mathcal{T}$ 
is defined by   the BRST cohomology as
\begin{align}
V_H(\mathcal{T}):=H^*(Q_{\rm BRST})=\mathrm{Ker} \, Q_{\rm BRST}/\mathrm{Im} \, Q_{\rm BRST} \,.
\end{align}
Here we give comments on properties of the BRST cohomology. The above construction of VOA also works for non-abelian gauge theories. 
In this case, the  current for the bc-ghost $J_{bc} $  enters  the definition of the BRST current.
Specifically, if a 3d $\mathcal{N}=4$ supersymmetric  gauge theory is the dimensional reduction of a 4d $\mathcal{N}=2$ superconformal gauge theory, the BRST charge is nilpotent  without introducing the fermion current. 
Then the definition of  VOAs for the dimensional reduction of the 4d  gauge theories are same as the one for the 4d superconformal gauge theory in \cite{Beem:2013sza}. 
For example, the VOA associated with 3d $\mathcal{N}=4$ $SU(2)$ gauge theory with four fundamental hypermultiplets is a current algebra  $\widehat{\mathfrak{so}}(8)_{-2}$.

\subsection{Relation with VOA associated with toric hyper-K\"{a}hler varieties}
\label{sec:hyperVOA}
In this section, we  point out the relation between the VOAs associated with 3d $\mathcal{N}=4$ abelian gauge theories defined in the pervious section and the VOAs  associated with   toric hyper-K\"{a}hler varieties in  \cite{MR4320811}.  
From the physics view point, a toric hyper-K\"{a}hler variety is the moduli space of the Higgs branch vacua of a 3d   $\mathcal{N}=4$ abelian gauge theory given by $\mu^{-1}_{\mathbb{R}}(0) \cap \mu^{-1}_{\mathbb{C}}(0)/ G$, where $\mu_{\mathbb{R}}$ is the real moment map 
associated with the D-term potential. 
The VOA in \cite{MR4320811} is defined by 
 a BRST cohomology associated with
 the current for symplectic bosons $(X_i,Y_i)$ and   Heisenberg vertex algebras, 
where the OPE of $(X_i,Y_i)$ is  same as \eqref{eq:OPEsb} in the previous section.
 The OPE of the Heisenberg vertex algebra is defined by  
\begin{align}
h_a(z) h_b(0) \sim  \frac{C_{a b}}{z^2},
\label{eq:HeiVA}
\end{align}
where $C_{ab}$ is defined by
\begin{align}
 C_{ab}:=\sum_{i=1}^L Q_{a, i} Q_{b, i}.
\end{align}
The BRST charge ${Q}^{\prime}_{\rm BRST}$ for the VOA associated with a toric hyper-K\"{a}hler variety is defined by 
\begin{align}
{Q}^{\prime}_{\rm BRST}= \sum_{a=1}^{L} \oint \frac{dz}{2 \pi {\rm i}} {\sf c}_a\left(   J^a_{\text{sb}}(z)+ h_a(z)\right).
\end{align}
Here the $J^a_{\text{sb}}$ and  ${\sf c}_a $ are  given by \eqref{eq:currentsb} and \eqref{eq:bcghost}, respectively.
Since the OPE of Heisenberg vertex algebra   \eqref{eq:HeiVA} has  the same  as  the fermion current $J_F$ with $\tilde{Q}_{a, i}={Q}_{a, i}$ in    \eqref{eq:fcuurent},
 the BRST charge ${Q}^{\prime}_{\rm BRST}$ is nilpotent. 
 Then the VOA associated with a toric hyperK\"ahler variety is defined by the BRST cohomology of ${Q}^{\prime}_{\rm BRST}$.
 Therefore we can regard 
 the VOAs associated with  3d  $\mathcal{N}=4$ H-twisted  abelian gauge theories   as  fermionic extensions of  the   VOAs associated with toric hyper-K\"{a}hler varieties   with the identification $h_a=J^a_{f}$:
\begin{align}
  V_H(\mathcal{T})=H^*(Q_{\rm BRST}) \supset H^*(Q^{\prime}_{\rm BRST}) |_{h_a \mapsto J^a_{f}}.
\end{align}

\section{Fermionic extensions of $W$-algebras and the mirror of  SQED}
\label{sec:section3}
In this section, we  consider a $G=\prod_{a=1}^{N-1}U(1)_a$  linear quiver gauge theory $\widetilde{\mathcal{T}}_{\mathrm{SQED}}^{ N}$ specified by the gauge charges
\begin{align}
Q_{a, i}=\delta_{a,i} -\delta_{a, i-1}\,,
\label{eq:mirrorQ}
\end{align}
with $i=1,\cdots,N$ and $a=1,\cdots,N-1$. The theory $\widetilde{\mathcal{T}}^{ N}_{\rm SQED}$ is the 3d mirror dual of $N$-flavors $U(1)$ SQED $\mathcal{T}_{\mathrm{SQED}}^{N}$ \cite{Intriligator:1996ex}.
The  moduli space of the Higgs branch vacua of $\widetilde{\mathcal{T}}_{\mathrm{SQED}}^{N}$ is an orbifold $\mathbb{C}^2/\mathbb{Z}_N$.
The currents of symplectic bosons  $(X_i, Y_i)$  and fermions $(\psi_i,\chi_i)$ associated with  $\widetilde{\mathcal{T}}^{ N}_{\rm SQED}$ are given by
\begin{align}
J^{i}_{\text{sb}}&=X_i Y_i- X_{i+1} Y_{i+1}, \\ 
J^{i}_{\text{f}}&=\psi_i \chi_i- \psi_{i+1} \chi_{i+1}\,,
\end{align}
with $i=1,\cdots,N-1$.
The OPE of the fermion current is given by
\begin{align}
J^i_{\text{f}} (z ) J^{j}_{\text{f}} (0) \sim \frac{C_{i j}}{z^2},
\label{eq:OPEFcurrent}
\end{align}
where $C_{i j}$ is the Cartan matrix for $\mathfrak{su}(N)$.

Motivated by the VOA associated with $\mathbb{C}^2 /\mathbb{Z}_N$,
 we define a total energy momentum tensor by 
\begin{align}
&T:= T_{\text{sb}}+T_{\text{f}}+T_{\text{bc}} \,,
\end{align}
where $ T_{\text{sb}}, T_{\text{f}} $ and $T_{\text{bc}}$ are defined by
\begin{align}
& T_{\text{sb}}:= \frac{1}{2} \sum_{i=1}^N \left(  X_i \partial Y_i- Y_i \partial X_i \right)\,, \\
&T_{\text{f}}:= \frac{1}{2} \sum_{i,j=1}^{N-1} C^{i j} J^{i}_{\text{f}} J^{j}_{\text{f}}  \,,
\label{eq:EMf}\\
&T_{\text{bc}}:= -\sum_{i=1}^{N-1}  {\sf b}_{i} \partial {\sf c}_{i}.
\end{align}
Here $C^{i j}$ is the inverse matrix of the Cartan matrix $C_{i j}$.  Then the total energy-momentum tensor $T$ becomes BRST closed by $Q_{\rm BRST}$.
Note that $T_{\rm f}$ is related to the canonical energy-momentum tensor of the fermions by 
\begin{align}
T_{\text{f}}+\frac{N}{2} J^2_F  =
\frac{1}{2} \sum_{i=1}^N \left( \partial \psi_i \cdot \chi_i -\psi_i \partial \chi_i  \right)\,.
\end{align}
Here $J_F$ is a BRST closed operator defined  below \eqref{eq:currentJF}.

The energy-momentum tensor ${T}^{\prime}$ in the VOA associated with $\mathbb{C}^2/{\mathbb{Z}_N}$ is  defined by
 the replacement $J^i_f \mapsto h^i$ in   \eqref{eq:EMf} as  
 \begin{align}
&{T}^{\prime}:= T_{\text{sb}}+T_{\text{h}}+T_{\text{bc}}, \quad  
T_{\text{h}}:= \frac{1}{2} \sum_{i,j=1}^{N-1} C^{i j} h_{i} h_{j}  . 
\end{align}
$T^{\prime}$ is  closed by  $Q^{\prime}_{\rm BRST}$.
We  define operators $G^{\pm}, J_{SB}$ by 
\begin{align}
&G^+ = \prod_{i=1}^{N}X_i , \quad G^- =  \prod_{i=1}^{N} Y_i ,  \quad J_{\text{SB}}:= \frac{-1}{N} \sum_{i=1}^N X_i Y_i.   
\end{align} 

Kuwabara \cite{MR4320811} has shown that the VOA $H^* (Q^{\prime}_{\rm BRST})$ associated with the orbifold $\mathbb{C}^2/{\mathbb{Z}_N}$ is isomorphic to 
a $W$-algebra $W^{-N+1}(\mathfrak{sl}_N,f_{\rm sub})$ with $N \ge 3$, which is defined by 
 a quantized Drinfeld-Sokolov reduction  by a sub-regular nilpotent element  $f_{\rm sub}$ of $\mathfrak{sl}_N$ with a level $k=-N+1$.
 Moreover $W^{-N+1}(\mathfrak{sl}_N,f_{\rm sub})$  is generated by operators $({T}^{\prime}, G^{\pm}, J_{SB})$. 
 From the discussion in  Section \ref{sec:hyperVOA}, 
 we conclude that the $W$-algebra $W^{-N+1}(\mathfrak{sl}_N,f_{\rm sub})$  is 
 a sub-algebra of  $V_H(\widetilde{\mathcal{T}}^N_{\rm SQED})$. In other words, 
 $V_H(\widetilde{\mathcal{T}}^N_{\rm SQED})$ is a fermionic extension of the $W$-algebra:
 \begin{align}
V_H(\widetilde{\mathcal{T}}^N_{\rm SQED}) \supset W^{-N+1}(\mathfrak{sl}_N,f_{\rm sub})\,.
\end{align} 

Since elements of the BRST cohomology are expected to be associated with gauge  invariant operators, we conjecture that the VOA 
$V_H(\widetilde{\mathcal{T}}^N_{\rm SQED})$ with $N\ge 3$
 is generated by operators $(J_{SB}, J_F, M^{\pm}_i, G^{\pm}_{I})$. Here $(J_F, M^{\pm}_i, G^{\pm}_{I})$ are defined by
\begin{align}
J_F&:=-\frac{1}{N}\sum_{i=1}^{N} \psi_i \chi_i\,,
\label{eq:currentJF}
\\
{G}^+_I &:=\prod_{i \in I^c}X_{i}  \prod_{j \in I} \chi_{j}\,, 
\label{eq:GpI} \\
G^{-}_I &:=\prod_{i \in I^c}Y_{i}  \prod_{j \in I} \psi_{j}\,,
\label{eq:GmI}
 \\
{M}^{+}_i &:=X_{i}   \psi_i \,, \\
{M}^{-}_i &:=Y_ i \chi_{i}\,,
\end{align} 
where $I=\{i_1,\cdots,i_{\ell} \}\subset \{1,\cdots, N \}$ with $|I|=\ell$, and $I^{c}$ is the complement of $I$ in $\{1,\cdots, N \}$.
For $N=3$, the total energy-momentum tensor is expressed by $(J_{SB}, J_F, M^{\pm}_i)$ as
\begin{align}
T =-\frac{1}{2} \left( 3 J_{SB}^2 +(J_{SB}+J_F)^2 +\sum_{i=1}^3 [M^{i}_+, M^{-}_i] \right)\,.
\label{eq:TJSBJF}
\end{align} 

Note that if we identify the symplectic bosons with the scalars $(q_i ,\tilde{q}_i)$ in the hypermultiplets and identify the complex fermions  $(\psi_i, \chi_i)$ with
the boundary  fermions in the $\mathcal{N}=(0,2)$ fermi multiplets,
${M}^{\pm}_i$ and ${G}^{\pm}_{I}$ are same as gauge invariant operators in $\widetilde{\mathcal{T}}^{N}_{\rm SQED}$ at the boundary.
In particular, BRST closed operators ${G}^{+}_{\o}=G^+ = \prod_{i=1}^N X_i$ and ${G}^{-}_{\o}={G}^{-} =\prod_{i=1}^N Y_i$ are identified  with 
baryons $q_1 q_2 \cdots q_N$ and $\tilde{q}_1 \tilde{q}_2 \cdots \tilde{q}_N$, respectively. 
On the other hand, operators $X_i Y_i$ , $i=1,\cdots, N$ associated the meson $q_1 \tilde{q}_1 = \cdots = q_N \tilde{q}_N$ are not BRST closed. An analog of the meson in the VOA is given  by
 \begin{align}
X_1 Y_1 +\psi_1 \chi_1=X_2 Y_2 +\psi_2 \chi_2= \cdots =X_N Y_N +\psi_N \chi_N=-(J_{SB}+ J_F)\,,
\end{align} 
where the equalities hold up to BRST exact terms.

In general, although it is  difficult to  determine  a generator of  BRST cohomologies by elementary methods,
we explicitly computed the OPE between  $(T,J_{SB}, J_F, M^{\pm}_i, G^{\pm}_{I})$ for $N=3$ and have shown that the OPE  is closed, which supports our conjecture.
The OPE of elements in the BRST cohomology   is written in appendix \ref{sec:appa}.   Note that  a sub-algebra generated by  $(T, G^{\pm}, J_{SB})$ in  $V_H (\widetilde{\mathcal{T}}^{3}_{\rm SQED})$ is a  Bershadsky-Polyakov algebra with the central charge $c=-5$.

\section{SUSY indices on $S^1 \times D^2$ and  the vacuum character of VOA}
\label{sec:section4}
In 4d/2d correspondence, the superconformal indices \cite{Romelsberger:2005eg, Kinney:2005ej} for 4d $\mathcal{N}=2$ SCFTs in the Schur limit 
\cite{Gadde:2011uv, Gadde:2011ik}
 conjecturally agree with the vacuum characters of the corresponding VOAs  \cite{Beem:2013sza}.
Similarly,  3d-2d coupled  supersymmetric indices $Z^{(H)}_{S^1 \times D^2}$ on $S^1 \times D^2$ with the H-twist conjecturally agree with the vacuum characters of the corresponding VOAs appearing at the boundary. 
When a 3d $\mathcal{N}=4$ gauge theory is a dimensional reduction of a 4d $\mathcal{N}=2$ superconformal gauge theory,
 the   index on $S^1 \times D^2$ with the H-twist agrees with the  Schur index. 
  For example, the H-twisted index $Z_{S^1 \times D^2}$ of the 3d $\mathcal{N}=4$ $SU(2)$ gauge theory with 
 four hypermultiplets in the fundamental representation is given by
 \begin{align}
Z^{(H), SU(2)+4\mathrm{hyp}}_{S^1 \times D^2}&=\frac{(q;q)^2_{\infty}}{2} \oint \frac{dx }{2 \pi {\rm i} x} \frac{(x^{2} ;q )^2_{\infty} (x^{-2} ;q )^2_{\infty} }
{(x q^{\frac{1}{2}} ;q )^8_{\infty} (x^{-1} q^{\frac{1}{2}};q )^8_{\infty}} \\
&=1 + 28 q + 329 q^2 + 2632q^3 + 16380 q^4 + \cdots \nonumber\,.
\end{align}
  The above expression is actually same as the 4d $\mathcal{N}=2$ superconformal index (for example see \cite{Dedushenko:2019yiw}) in the Schur limit and agrees with the expansion of 
  the vacuum character of an affine current algebra $\widehat{\mathfrak{so}}(8)_{-2}$ \cite{Beem:2013sza}.

This observation suggests that the vacuum character of   $V_H(\widetilde{\mathcal{T}})$ is given by
the 3d  H-twisted index $Z^{(H), \widetilde{\mathcal{T}}}_{ S^1 \times D^2}$ of  3d theory $\widetilde{\mathcal{T}}$ on $ S^1 \times D^2$  including   the elliptic genus of 
2d $\mathcal{N}=(0,2)$ fermi multiplets at the boundary torus. 
Moreover, in  3d $\mathcal{N}=4$ mirror symmetry,   the C-twisted index $Z^{(C), \widetilde{\mathcal{T}}}_{S^1 \times D^2}$ of the theory $\mathcal{T}$ mirror dual to $\widetilde{\mathcal{T}}$  is expected to agree with  the index $Z^{(H), {\widetilde{\mathcal{T}}}}_{S^1 \times D^2}$, where  two theories:  $\mathcal{T}$ with 2d $\mathcal{N}=(0,2)$ fermi multiplets  and $\widetilde{\mathcal{T}}$ have to satisfy the 't Hooft anomaly matching condition \cite{Dimofte:2017tpi}.

Let us compare the supersymmetric indices on $ S^1 \times D^2$ and  elements of  BRST cohomology at low dimensions.
First we fix the charge assignments of the mirror pair by the anomaly matching condition. To see anomaly matching, it is convenient to
introduce anomaly polynomials \cite{Dimofte:2017tpi}.
The anomaly polynomial of 3d theory $\widetilde{\cal T}^{N}_{\rm SQED}$ with H-twist is given by
\begin{align}
\mathcal{I}^{(H)}_{\rm mirror}&=- (\mathbf{f}_1-\mathbf{y}_1)^2 -\sum_{a=2}^{N-2} (\mathbf{f}_{a}-\mathbf{f}_{a-1})^2 - (-\mathbf{f}_{N-1}+\mathbf{y}_2)^2  \,.
\end{align}
Here $\mathbf{f}_a$ is the field strength of  the $U(1)_a$ gauge group with  $a=1,\cdots, N-1$.  $\mathbf{y}_1= - \mathbf{y}_2=2\mathbf{y}$  is the field strength of  the background gauge field of $U(1)_y$ flavor symmetry acting
on   the first and the $N$-th hypermultiplets as $(q_1,\tilde{q}_1) \mapsto (e^{-{\rm i} \frac{\theta}{2}} q_1, e^{{\rm i} \frac{\theta}{2}}\tilde{q}_1)$ and $(q_N,\tilde{q}_N) \mapsto (e^{-{\rm i} \frac{\theta}{2}} q_N, e^{{\rm i} \frac{\theta}{2}}\tilde{q}_N)$.
The boundary fermi multiplets  contribute to the anomaly polynomial as
\begin{align}
\mathcal{I}_{\rm fermi}&=  (-\mathbf{f}_1+\mathbf{y}_1+\mathbf{s})^2+\sum_{i=2}^{N-2} (-\mathbf{f}_{a}+\mathbf{f}_{a-1}+s)^2+   (-\mathbf{y}_2+\mathbf{f}_{N-1}+\mathbf{s})^2 \,.
\end{align}
Here $\mathbf{s}$ is the background field strength of $U(1)_s$ flavor symmetric acting on fermi multiplets as $(\chi_i, \psi_i) \mapsto  (e^{{\rm i} \theta} \chi_i,e^{-{\rm i} \theta} \psi_i)$. 
The anomaly polynomial of the 3d-2d system is given by
\begin{align}
\mathcal{I}^{(H)}_{\rm mirror}+\mathcal{I}_{\rm fermi}&=N \mathbf{s}^2+2 \mathbf{s} (\mathbf{y}_2- \mathbf{y}_1)
\end{align}
Since the $\mathbf{f}_i$'s cancel out between the 3d and 2d fermions, the gauge anomaly is absent.

In the C-twisted $N$-flavors SQED,  the $U(1)$ gauge symmetry at the boundary survives as a global symmetry, which is identified with 
$U(1)_s$-symmetry in the mirror side. The topological $U(1)$ symmetry acting on Coulomb branch operators is identified with the flavor 
$U(1)_y$ symmetry in the mirror side. Then the anomaly polynomial of C-twisted $N$-flavors SQED is given by
\begin{align}
\mathcal{I}^{(C)}_{\rm SQED}&=  N \mathbf{s}^2  -2 \mathbf{s} \mathbf{y}\,.
\label{eq:anomalyC}
\end{align}
Here the first term in  \eqref{eq:anomalyC} is the effective $U(1)_s$ CS-term generated by $N$-hypermultiplets with the Dirichlet boundary condition.
The second term is the $U(1)_s$-$U(1)_y$ mixed CS-term. We find the agreement $\mathcal{I}^{(H)}_{\rm mirror}+\mathcal{I}_{\rm fermi}=\mathcal{I}^{(C)}_{\rm SQED}$.

Now we write down formulas of indices on $S^1 \times D^2$ of the mirror pair. 
The supersymmetric index of a 3d $\mathcal{N} \ge 2$ theory on $S^1 \times D^2$ is defined by 
\begin{align}
Z_{S^1 \times D^2} = \mathrm{Tr}_{\mathcal{H}(D^2)} (-1)^F q^{J_3+\frac{R}{2}} \prod_{f} y^{F_f}_f
\label{eq:indS1xD2}
\end{align}
Here $\mathcal{H}(D^2)$ is the Hilbert space on 2d hemisphere $D^2$ with  appropriate boundary conditions for the fields at the boundary $\partial (S^1 \times D^2)=\mathbb{T}^2$. $F$, $J_3$, $R$ and $F_f$ are  the fermion number, the rotation of $D^2$, an R-charge and the flavor charges of the 3d $\mathcal{N}$=2 theory.
The parameters $q$ and $w_f$ are fugacities for these charges.  The indices have been first studied in \cite{Yoshida:2014ssa} in detail for  the Neumann boundary condition of the $\mathcal{N}=2$ vector multiplet, and later studied in \cite{Dimofte:2017tpi} by operator counting arguments.  The indices with the Dirichlet boundary condition for the $\mathcal{N}=2$ vector multiplet  has been proposed in \cite{Dimofte:2017tpi}, and later derived by supersymmetric localization in \cite{Bullimore:2020jdq}.
From  formulas of $Z_{S^1 \times D^2} $ in \cite{Yoshida:2014ssa, Dimofte:2017tpi, Bullimore:2020jdq},
the H/C-twisted indices of $\prod_{a=1}^{N-1}U(1)_a$ linear quiver gauge theory and its mirror dual i.e., $U(1)$ $N$-flavors SQED
 are written as
\begin{align}
Z_{S^1 \times D^2}^{ (H),\widetilde{\mathcal{T}}^N_{\rm SQED}}&=(q;q)^{2(N-1)}_{\infty} 
\oint \prod_{i=1}^{N-1} \frac{dx_i}{2 \pi {\rm i} x_i} \prod_{i=1}^{N} \frac{ (x^{-1}_{i} x_{i-1} s q^{\frac{1}{2} } ;q )_{\infty} (x_{i} x^{-1}_{i-1} s^{-1} q^{\frac{1}{2} };q )_{\infty}}{ ( x_{i} x^{-1}_{i-1}  q^{\frac{1}{2}};q)_{\infty} ( x^{-1}_{i} x_{i-1} q^{\frac{1}{2}};q)_{\infty}} 
\label{eq:Hindexmirror}
\,, \\
Z_{S^1 \times D^2}^{(C), \mathcal{T}^N_{\rm SQED}}
&=\frac{1}{(q;q)^2_{\infty} }  \sum_{m \in  \mathbb{Z}} q^{ N m^2 } 
(-s)^{N m} y^m
  ( s  q^{1+m };q )^N_{\infty}  ( s^{-1}  q^{1-m};q )^N_{\infty} \,.
\end{align}
Here $y^{\frac{1}{2}}= x_0=x_N^{-1}$ is the  fugacity of the $U(1)_y$ flavor symmetry. The numerator of the integrand of \eqref{eq:Hindexmirror} is 
the elliptic genus of the boundary $\mathcal{N}=(0,2)$ fermi multiplets.
 
The 3d mirror symmetry predicts that the two indices of the mirror pair agree each other: 
\begin{align}
Z_{S^1 \times D^2}^{ (H),\widetilde{\mathcal{T}}^N_{\rm SQED}}=Z_{S^1 \times D^2}^{(C), \mathcal{T}^N_{\rm SQED}}. 
\label{eq:mirrorind}
\end{align}
We can check \eqref{eq:mirrorind} in an order-by-order computation of  $q^{\frac{1}{2}}$.
For example, $N=3$ case:
\begin{align}
&Z_{S^1 \times D^2}^{ (H),\widetilde{\mathcal{T}}^3_{\rm SQED}}=Z_{S^1 \times D^2}^{(C), \mathcal{T}^3_{\rm SQED}}=
\nonumber \\
&=1+\left[ 2 -3\left(s+ \frac{1}{s} \right)  \right]q+[ \left(y+\frac{1}{y} \right) -3 \left(y s + \frac{1}{y s } \right)
+3 \left(y s^2 + \frac{1}{y s^2} \right)-\left( y s^3 + \frac{1}{y s^3} \right)   ] q^{\frac{3}{2}} \nonumber \\
&+ \left[ 14-6 \left(s+\frac{1}{s} \right)+3\left(s^2+\frac{1}{s^2} \right)
  \right] q^2+O(q^{\frac{5}{2}})
\end{align}

In order to relate the indices on $S^1 \times D^2$ with the VOA, we define a canonical total energy-momentum tensor $\tilde{T}$ by
\begin{align}
\tilde{T}:=T+\frac{N}{2} J_F J_F={T}_{sb}+T_{bc}+\frac{1}{2} \sum_{i=1}^N \left( \partial \psi_i \cdot \chi_i -\psi_i \partial \chi_i  \right)\,.
\end{align}
Then the canonical energy-momentum tensor $\tilde{T}$ and $\mathcal{O} \in \{ J_{SB}, J_F, M^{\pm}_i, G^{\pm}_{I} \}$ 
 satisfy the following OPE:
\begin{align}
\tilde{T}(z) \mathcal{O}(0)  \sim 
\frac{h_{\mathcal{O}} \mathcal{O} }{z^2}  +\frac{\partial \mathcal{O} }{z} .
\end{align}
Here the  dimensions of  primary fields are $( h_{J_{SB}}, h_{J_F}, h_{M^{\pm}_i}, h_{G^{\pm}_{I}} )=(1, 1, N/2, N/2)$.
We define  $L_n $, $J_{F,n}$ and $J_{B,n}$ by
\begin{align}
L_n&=  \oint \frac{dz}{2 \pi {\rm i} z} z^{n+2}\, \tilde{T}(z)\,, \\  
J_{F,n}&= 3 \oint \frac{dz}{2 \pi {\rm i} z}    z^{n+1}J_{F}(z) \,, \\
 J_{B,n}&=\oint \frac{dz}{2 \pi {\rm i} z}    z^{n+1} (J_{SB}(z)+J_F(z))\,,
\end{align}
where $n \in \mathbb{Z}$.
Note that $J_{ F, 0}$ is the fermion number operator which counts the number of  $\chi_i$  minus the number of $\psi_i$.
A charge  $J_{B,0}$ is   thought as an analog of the $U(1)_y$-flavor symmetry in the VOA.
We consider a refinement of the vacuum character $\mathrm{Tr} (-1)^{J_{F,0}} q^{L_0}$ of $V_H(\widetilde{\mathcal{T}}^N_{\mathrm{SQED}})$   by $J_{F,0}$, $J_{B,0}$:
\begin{align}
\chi_{V_H(\widetilde{\mathcal{T}}^N_{\rm SQED})} (q,y,s)=\mathrm{Tr} (-1)^{J_F,0} q^{L_0} y^{J_{B, 0}} s^{J_{F, 0}} \,.
\end{align}
 We conjecture that the character $\chi_{V_H(\widetilde{\mathcal{T}}^N_{\rm SQED})}$
  agrees with the H-twisted index of $\widetilde{\mathcal{T}}^N_{\rm SQED}$:  $Z_{S^1 \times D^2}^{(H), \widetilde{\mathcal{T}}^N_{\rm SQED}}$ and the C-twisted index of $\mathcal{T}^N_{\rm SQED}$: $Z_{S^1 \times D^2}^{(C), \mathcal{T}^N_{\rm SQED}}$, namely
  \begin{align}
\chi_{V_H(\widetilde{\mathcal{T}}^N_{\rm SQED})}=Z_{S^1 \times D^2}^{ (H),\widetilde{\mathcal{T}}^N_{\rm SQED}}
=Z_{S^1 \times D^2}^{(C), \mathcal{T}^N_{\rm SQED}}\,.
\end{align}
 The series expansions of $Z_{S^1 \times D^2}$ in   the various order of $q^{\frac{1}{2}}$ and $N=3,4,5,6,\cdots,$  suggest that
 the indices are given by
 \\
\underline{$N$:even number} 
\begin{align}
 Z_{S^1 \times D^2}^{ (H),\widetilde{\mathcal{T}}^{N={\rm even}}_{\rm SQED}} 
 =Z_{S^1 \times D^2}^{(C), \mathcal{T}^{N={\rm even}}_{\rm SQED}}&
 =1+
  \sum_{n=2}^{\infty}  \left\{
\begin{array}{ll}
 c_{\frac{n}{2}}   q^{\frac{n}{2}} & ( n= { \rm even}  )\\
 0 & ( n={\rm odd}  ) \\
\end{array}
\right\}\,.
\label{eq:Indexpand}
\end{align}
 \\
\underline{$N$:odd number } 
{\small{
\begin{align}
 Z_{S^1 \times D^2}^{ (H),\widetilde{\mathcal{T}}^{N={\rm odd}}_{\rm SQED}}&=Z_{S^1 \times D^2}^{(C), \mathcal{T}^{N={\rm odd}}_{\rm SQED}} 
 =1
 +\sum_{n=2}^{N} c_{\frac{n}{2}} q^{\frac{n}{2}} 
 \nonumber
\\
 &
+ \sum_{n=N+1}^{2N-1}  \left\{
\begin{array}{ll}
 c_{\frac{n}{2}} (y, s)  q^{\frac{n}{2}} & (    n={\rm odd})\\
 c_{\frac{n}{2}} (s) q^{\frac{n}{2}}& (   n={\rm even}  ) \\
\end{array}
\right\}\,
+ \sum_{n=2 N}^{\infty} c_{\frac{n}{2}}  q^{\frac{n}{2}}\,.
 \label{eq:Indexpand2}
 \end{align}
}}
Here $\sum_{n=2}^{N}c_{\frac{n}{2}}q^{\frac{n}{2}}$ agrees with the  $q^{\frac{1}{2}}$-expansion of the plethystic exponential 
of the single letter index $f(y,s,q)$ of $J_{SB}, J_F, M^{\pm}_i, G^{\pm}_{I}$ with $\partial$'s:
 \begin{align}
c_{\frac{n}{2}} q^{\frac{n}{2}}=\exp\left(\sum_{n=1}^{\infty} \frac{f(y^n, s^n, q^n)}{n} \right) \Big|_{q^{\frac{n}{2}}}, \quad (2 \le n \le N)\,.
\end{align}
with 
\begin{align}
f(y,s,q) &=\frac{1}{1-q}\left[ \left(2 -N\left(s+\frac{1}{s} \right)\right) q+\left( y+ \frac{1}{y}+\sum_{l=-N  \atop l \neq 0}^{N} (-1)^l  y^{{\rm sign}(l)} \binom{N}{l}  s^l \right) q^{\frac{N}{2}}
 \right]\,.
 \end{align}

 The expansion of the H/C-twisted indices has the following properties on the VOA.
First,  $\sum_{n=2}^{N}c_{\frac{n}{2}} q^{\frac{n}{2}}$  in \eqref{eq:Indexpand} and \eqref{eq:Indexpand2}
 agree with   in the plethystic exponential of the single letter index of $J_{SB}, J_F, M^{\pm}_i, G^{\pm}_{I}$.
 This agreement suggests that any operator  $\mathcal{O}$  with  ${\rm dim} \, \mathcal{O} \le N/2$ is freely generated by $J_{SB}, J_F, M^{\pm}_i, G^{\pm}_{I}$. On the other hand, higher order terms in the H/C-twisted index do not coincide with the the plethystic exponential, which imply  null operators should exist.
 When $N={\rm even}$, the half-integer power of $q$ does not appear in the expansion. This suggests that there are no operators consisting of an odd number of $X_i, Y_i, \chi_i,$ and $ \psi_i$. 
For $N={\rm odd}$, the  coefficients of the first term in the second line of \eqref{eq:Indexpand2} 
are consistent with the charges and dimensions of operators consisting
of  $(J_{SB}, J_F, M^{\pm}_i, G_{I})$. 
These observations may provide an evidence that the VOA  $V_H(\widetilde{\mathcal{T}}^{N}_{\rm SQED})$ is generated by operators 
$(J_{SB}, J_F, M^{\pm}_i, G_{I})$,  and that the H/C-twisted indices are the vacuum character of the VOA.

\rem{
\begin{align}
Z_{S^1 \times D^2}^{ (H),\widetilde{\mathcal{T}}^N_{\rm SQED}}
&=1+\left[\right]q
&+\sum{l =1 }^{\infty} a_n q^{ l}
\end{align}
}


\section{Summary and discussion}
\label{sec;section5}
In this paper, we have studied properties of the vertex operator algebras  associated with 3d $\mathcal{N}=4$ abelian gauge theories. We have pointed out that  these VOAs are fermionic extensions of the VOAs associated with toric hyper-K\"ahler varieties. Specifically, we have focused on the 3d mirror of $N$-flavors $U(1)$ SQED and analyzed the structure of the VOA, which is a fermionic extension of a $W$-algebra $W^{-N+1}(\mathfrak{sl}_N, f_{\rm sub})$.  For $N$=3, we have shown that the OPE of the generator are closed. 
We conjectured that  the VOA is generated by $(J_{SB}, J_F, M^{\pm}_i, G^{\pm}_I)$ and proposed expressions for the vacuum character of the VOA in terms of the H/C-twisted indices. Additionally, we have studied the relationship between the generator and the H/C-twisted indices of the 3d ${\cal N}=4$ mirror pair.

We comment on future directions of our research.
In recent years, VOAs associated  with 3d $\mathcal{N}\ge2$ theories  have been studied in \cite{ Cheng:2018vpl, Costello:2020ndc, Zeng:2021zef, Ballin:2022rto, Garner:2022rwe, Dedushenko:2022fmc, Alekseev:2022gnr}. 
On an interval $\mathbb{R}^2 \times I$, VOAs living on the boundaries are enriched by introducing line operators \cite{Alekseev:2022gnr}.
Characters of VOAs associated 3d $\mathcal{N}=2$ theories  have interesting modular properties  such as mock modular form \cite{Cheng:2018vpl}. It is interesting to study these properties in the fermionic extension of the $W$-algebra.

 In this paper we have studied properties of the VOA $V_H(\widetilde{\cal T}^N_{\rm SQED})$ by elementary methods.
  There are many important examples of VOAs  obtained by quantized Drinfeld-Sokolov reduction of Lie algebra, such as superconformal algebras and   $W$-algebras.  Since the VOA $V_H(\widetilde{\cal T}^N_{\rm SQED})$ contains a $W$-algebra, we expect that the VOA $V_H(\widetilde{\cal T}^N_{\rm SQED})$  itself can be obtained by a quantized Drinfeld-Sokolov reduction of a Lie superalgebra.

 As mentioned in  previous sections, a 4d superconformal gauge theory and its 3d reduction have the same VOA. On the other hand 
the situation is more subtle  for non-Lagrangian cases. For example,  let us consider $(A_1, A_3)$ Argyres-Douglas(AD) theory which is a  non-Lagrangian theory.  The 3d reduction of $(A_1, A_3)$ AD theory is the $T[SU(2)]$ \cite{Buican:2015hsa},  same as $2$-flavor SQED and its mirror dual: ${\cal T}^{2}_{\rm SQED} \simeq \widetilde{{\cal T}}^{2}_{\rm SQED}$.  
In our analysis, the VOA $V_H (\widetilde{\mathcal{T}}^{2}_{\rm SQED})$ contains Grassmann odd generators $X_i \psi_i, Y_i \chi_i, X_i \chi_j,$ and $Y_i \psi_j$ with $1 \le i \neq j \le 2$, (see appendix). On the other hand,  the VOA for $(A_1, A_3)$ AD theory is an affine current algebra $\widehat{\mathfrak{su}}(2)_{-\frac{4}{3}}$  \cite{Buican:2015ina}. In this case,  
VOAs associated with 4d $\mathcal{N}=2$ non-Lagrangian SCFTs and their 3d reductions appear to be unrelated to each other. It would be interesting to study VOAs associated with 4d non-Lagrangian theories with a boundary and to study the 3d reductions of them. 

Another interesting direction is to study VOAs associated with the C-twist. Although a VOA associated with the C-twist is believed to agree with the VOA associated with its mirror dual with the H-twist,
it is difficult to analyze elements in VOAs with the C-twist associated with boundary monopole operators. Since baryons are dual to monopole operators, $G^{\pm}$ should be dual to  elements associated with two monopole operators with minimal charges. It is interesting to study monopole operators in the VOA associated with the C-twisted $N$-flavor SQED and show the isomorphism between the VOAs associated with the H/C-twisted mirror pair.

\section*{Acknowledgements}
The author would like to thank Tomoyuki Arakawa, Hiraku Nakajima, and Takahiro Nishinaka for their discussions. 
He would also like to thank Mykola Dedushenko for his comments.
This work is supported by Grant-in-Aid for Scientific Research 21K03382, JSPS.

\appendix

\section{OPE of  elements in  BRST cohomologies }
\label{sec:appa}
In this appendix we calculate OPE of elements of  BRST cohomologies for $N=2,3$ .
For $N=3$, in order to compactly express the OPE, we define $ F^{\pm}_i,G^{\pm}_i, i=1,2,3$ and $F^{\pm}$  by
\begin{align}
&F^+=\prod_{i=1}^{3} \chi_i:, \quad F^- =  \prod_{i=1}^{3} \psi_i, 
\nonumber \\
& F^{+}_1:=  X_3 X_2 \chi_1, \quad F^{+}_2:=  X_3 X_1 \chi_2, \quad F^{+}_3:=  X_1 X_2 \chi_3, \nonumber \\ 
& F^{-}_1:=  Y_3 Y_2 \psi_1, \quad F^{-}_2:=  Y_3 Y_1 \psi_2, \quad F^{-}_3:=  Y_1 Y_2 \psi_3, \nonumber \\ 
& G^{+}_1:=   X_1 \chi_2 \chi_3 , \quad G^{+}_2:=  X_2 \chi_3  \chi_1, \quad G^{+}_3:=X_3  \chi_1 \chi_2 , \nonumber \\ 
& G^{-}_1:= Y_1 \psi_2  \psi_3, \quad G^{-}_2:=Y_2  \psi_3 \psi_1 , \quad G^{-}_3:=Y_3  \psi_1 \psi_2 . 
\end{align}
Note that $G^{\pm}_i,  F^{\pm}_i , i=1,2,3$ and $F^{\pm}$ are same as \eqref{eq:GpI} and  \eqref{eq:GmI} up to the sign factor. 
The OPE of $(T, J_{SB}, G^{\pm}, J_F, F^{\pm},M^{\pm}_i, F^{\pm}_i, G^{\pm}_i ), i=1, 2, 3$ are given by
{\footnotesize{
\begin{align}
T(z)  T(0) &\sim \frac{-5}{2 z^4} +\frac{2 T}{z^2}+\frac{\partial T}{z}\,, \\
T(z)  J_{SB}(0) &\sim \frac{J_{SB}}{z^2} +\frac{\partial J_{SB}}{z}\,, \\
T(z)  G^{\pm}(0) &\sim \frac{3G^{\pm}}{2z^2} +\frac{\partial G^{\pm}}{z}\,, \\
T(z)  J_{F}(0) &\sim 0\,, \\
T(z)  F^{\pm}(0) &\sim 0\,, \\
T(z) M^{\pm}_i (0) &\sim \frac{5 M^{\pm}_i }{6 z^2} +\frac{1}{z} \left(\pm J_{F} M^{\pm}_i  + \partial M^{\pm}_i  \right)\,,  \\
T(z) F^{\pm}_i(0) &\sim \frac{4 F^{\pm}_i}{3 z^2}+  \frac{1}{z}\left(\frac{-1}{3} \varepsilon_{i j k } M^{\pm}_j G^{\pm}_k+\frac{2}{3} 
\partial F^{\pm}_i\right)\,, \\
T(z) G^{\pm}_i(0) &\sim \frac{5 G^{\pm}_i}{6 z^2}+ \frac{1}{z} \left(\frac{2}{3}M^{\pm}_i F^{\pm} +\frac{1}{3} 
\partial G^{\pm}_i\right)\,, \\
J_{SB} (z) J_{SB} (0)  &\sim \frac{-1}{3 z^2}\,, \\
J_{SB} (z) G^{\pm} (0) &\sim \frac{\pm G^{\pm} }{z}\,, \\
J_{SB}(z) J_F (0) &\sim 0\,, \\
J_{SB}(z) F^{\pm}(0) &\sim 0\,,\\
J_{SB}(z) M^{\pm}_i(0) &\sim  \frac{\pm M^{\pm}_i}{3 z}\,, \\
J_{SB}(z) F^{\pm}_i(0) &\sim  \frac{\pm 2 F^{\pm}_i}{3 z}\,, \\
J_{SB}(z) G^{\pm}_i(0) &\sim  \frac{\pm G^{\pm}_i}{3 z}\,, \\
G^{\pm} (z) G^{\pm} (0) &\sim 0\,, \\
G^{+} (z) G^- (0) &\sim \frac{1}{z^3}+\frac{-3 J_{SB} }{z^2} + \frac{1}{z} \left( 3 J_{SB}^2-\frac{3}{2} \partial J_{SB}-T \right)\,, \\
G^{\pm} (z) J_F (0) &\sim 0\,, \\
G^{\pm} (z) F^{\pm} (0) &\sim 0\,, \\
G^{\pm} (z) F^{\mp} (0) &\sim 0\,,  \\
G^{\pm} (z) M^{\pm}_i (0) &\sim 0\,, \\
G^{\pm} (z) M^{\mp}_i (0) &\sim \pm \frac{F^{\pm}_i   }{z}\,, \\
G^{\pm} (z) F^{\pm}_i (0) &\sim 0\,, \\
G^{\pm} (z) F^{\mp}_i (0) &\sim \frac{M^{\pm}_i   }{z^2} + \frac{1}{z} ( \mp 2 J_{SB} M^{\pm}_i  \pm J_{F} M^{\pm}_i  + \partial M^{\pm}_i)\,, \\
G^{\pm} (z) G^{\pm}_i(0)  &\sim 0\,, \\
G^{\pm} (z) G^{\mp}_i (0) &\sim \pm \frac{ \varepsilon_{i j k}  M^{\pm}_i M^{\pm}_j  }{2 z}\,, \\
J_{F} (z) J_{F} (0)  &\sim \frac{1}{3 z^2}\,, \\
J_{F} (z) F^{\pm} (0) &\sim \frac{\pm F^{\pm}  }{z}\,, \\
J_{F} (z) M^{\pm}_i (0)  &\sim  \frac{\mp M^{\pm}_i}{3z}\,, \\
J_{F} (z) F^{\pm}_i (0)  &\sim  \frac{\pm F^{\pm}_i}{3 z}\,, \\
J_{F} (z) G^{\pm}_i (0)  &\sim  \frac{\pm 2 G^{\pm}_i}{3 z}\,, \\
F^{\pm}(z)  F^{\pm}(0) &\sim 0\,, \quad \\ 
F^{+} (z) F^- (0) &\sim \frac{-1}{z^3}+\frac{-3 J_{F} }{z^2} - \frac{1}{z} \left( \frac{9}{2} J_{F}^2+\frac{3}{2} \partial J_{F} \right)\,, \\
 F^{\pm} (z) M^{\pm}_i (0) &\sim \frac{G^{\pm}_i}{z}\,, \\
 F^{\pm} (z) M^{\mp}_i (0) &\sim 0\,, \\
   F^{\pm} (z) F^{\pm}_i (0) &\sim 0\,, \\
F^{\pm} (z) F^{\mp}_i (0) &\sim \frac{ \varepsilon_{i j k}  M^{\mp}_j M^{\mp}_k  }{2z} \,, \\
F^{\pm} (z) G^{\pm}_i (0) &\sim 0\,,  \\
F^{\pm} (z) G^{\mp}_i (0) &\sim \frac{- M^{\mp}_i }{z^2}+\frac{\mp 3 J_F M^{\mp}_i  }{z}
\\
M^{\pm}_i (z) M^{\pm}_j (0) &\sim 0, \\
M^+_i (z) M^-_j (0) &\sim \frac{\delta_{ i j}}{z^2}- \frac{\delta_{ i j}}{z} (J_{SB}+J_F)\,, \\
M^{\pm}_i (z) F^{\pm}_j (0) &\sim \delta_{i j} \frac{G^{\pm} }{z}  \\
M^{\pm}_i (z) F^{\mp}_j (0) &\sim \frac{ \pm \varepsilon_{i j k} G^{\mp}_k }{z}\,, \\
  M^{\pm}_i (z) G^{\pm}_j (0) &\sim \frac{ - \varepsilon_{i j k} F^{\pm}_k }{z}\,, \\
M^{\pm}_i (z) G^{\mp}_j (0) &\sim \pm \delta_{i j} \frac{F^{\mp} }{z}\,,   \\
F^{\pm}_i (z) F^{\pm}_j (0) &\sim 0\,,   \\
F^{+}_i (z) F^{-}_j (0) &\sim  \frac{M^{-}_i M^{+}_j}{z}, \quad ( i \neq j),   \\
F^{+}_i (z) F^{-}_i (0) &\sim \frac{1}{z^3}+\frac{1}{z^2}(J_F-2 J_{SB})+\frac{1}{z} \left(J^2_{SB}+J^2_{F} -J_{SB}J_F -M^{+}_i M^{-}_i-\frac{3}{2} \partial J_{SB}  -T  \right), \\
F^{\pm}_i (z) G^{\pm}_i (0) &\sim 0 \,, \\
F^{\pm}_i (z) G^{\mp}_j (0) &\sim \mp\frac{\varepsilon_{i j k}M^{\pm}_k}{z^2}- \frac{1}{z}\varepsilon_{i j k}M^{\pm}_k (2 J_F -J_{SB}) \,, \\
G^{\pm}_i (z) G^{\pm}_j (0) &\sim 0\,, \\
G^{+}_i (z) G^{-}_j (0) &\sim - \frac{M^{+}_i M^{-}_j}{z}, \quad ( i \neq j)\,,   \\
G^{+}_i (z) G^{-}_i (0) &\sim \frac{-1}{z^3}+\frac{1}{z^2}(J_{SB}-2 J_F)+\frac{1}{z} \left(-\frac{3}{2} J_F^2+3 J_F J_{SB} -M^{+}_i M^{-}_i-\frac{3}{2} \partial J_F \right)\,. 
\end{align}
}}

For $N=2$, in order to  express the OPE, we define $ F^{\pm}_i, i=1,2$ and $F^{\pm}$  by
\begin{align}
&F^+= \chi_1 \chi_2:, \quad F^- =   \psi_1 \psi_2, 
\nonumber \\
& F^{+}_1:=  X_2  \chi_1, \quad F^{+}_2:=  X_1 \chi_2,  \nonumber \\ 
& F^{-}_1:=   Y_2 \psi_1, \quad F^{-}_2:=  Y_1 \psi_2, \
, \nonumber \\ 
\end{align}
Then  OPE of  elements for $2$-flavors SQED ($T[SU(2)]$ theory) is given by
{\footnotesize{
\begin{align}
T(z)  T(0) &\sim \frac{-3}{2 z^4} +\frac{2 T}{z^2}+\frac{\partial T}{z}\,, \\
T(z)  J_{SB}(0) &\sim \frac{J_{SB}}{z^2} +\frac{\partial J_{SB}}{z}\,, \\
T(z)  G^{\pm}(0) &\sim \frac{G^{\pm}}{z^2} +\frac{\partial G^{\pm}}{z}\,, \\
T(z)  J_{F}(0) &\sim 0\,, \\
T(z)  F^{\pm}(0) &\sim 0\,, \\
T(z) M^{\pm}_i (0) &\sim \frac{3 M^{\pm}_i }{4 z^2} +\frac{1}{z} 
\left(\pm J_{F} M^{\pm}_i  + \partial M^{\pm}_i  \right)\,,  \\
T(z) F^{\pm}_i(0) &\sim \frac{3 F^{\pm}_i}{4 z^2}+  \frac{1}{2 z}\left( - \varepsilon_{i j} M^{\pm}_j F^{\pm} + \partial F^{\pm}_i\right)\,, \\
J_{SB} (z) J_{SB} (0)  &\sim \frac{-1}{2 z^2}\,, \\
J_{SB} (z) G^{\pm} (0) &\sim \frac{\pm G^{\pm} }{z}\,, \\
J_{SB}(z) J_F (0) &\sim 0\,, \\
J_{SB}(z) F^{\pm}(0) &\sim 0\,,\\
J_{SB}(z) M^{\pm}_i(0) &\sim\pm  \frac{ M^{\pm}_i}{2 z}\,, \\
J_{SB}(z) F^{\pm}_i(0) &\sim  \pm \frac{ F^{\pm}_i}{2 z}\,, \\
G^{\pm} (z) G^{\pm} (0) &\sim 0\,, \\
G^{+} (z) G^- (0) &\sim \frac{1}{z^2}+\frac{-2 J_{SB} }{z} \,, \\
G^{\pm} (z) J_F (0) &\sim 0\,, \\
G^{\pm} (z) F^{\pm} (0) &\sim 0\,, \\
G^{\pm} (z) F^{\mp} (0) &\sim 0\,,  \\
G^{\pm} (z) M^{\pm}_i (0) &\sim 0\,, \\
G^{\pm} (z) M^{\mp}_{i} (0) &\sim \pm \frac{F^{\pm}_{i}   }{z}\,, \\
G^{\pm} (z) F^{\pm}_i (0) &\sim 0\,, \\
G^{\pm} (z) F^{\mp}_i (0) &\sim \pm \frac{M^{\pm}_i}{z}\,, \\
J_{F} (z) J_{F} (0)  &\sim \frac{1}{2 z^2}\,, \\
J_{F} (z) F^{\pm} (0) &\sim \frac{\pm F^{\pm}  }{z}\,, \\
J_{F} (z) M^{\pm}_i (0)  &\sim  \frac{\mp M^{\pm}_i}{2z}\,, \\
J_{F} (z) F^{\pm}_i (0)  &\sim  \frac{\pm F^{\pm}_i}{2 z}\,, \\
F^{\pm}(z)  F^{\pm}(0) &\sim 0\,, \quad \\ 
F^{+} (z) F^- (0) &\sim \frac{-1}{z^2}+\frac{-2 J_{F} }{z}
\,, \\
 F^{\pm} (z) M^{\pm}_i (0) &\sim -\frac{ \varepsilon_{i j} F^{\pm}_i}{z}\,, \\
 F^{\pm} (z) M^{\mp}_i (0) &\sim 0\,, \\
   F^{\pm} (z) F^{\pm}_i (0) &\sim 0\,, \\
F^{\pm} (z) F^{\mp}_i (0) &\sim -\frac{ \varepsilon_{i j }  M^{\mp}_j   }{z} \,, \\
M^{\pm}_i (z) M^{\pm}_j (0) &\sim 0, \\
M^+_i (z) M^-_j (0) &\sim \frac{\delta_{ i j}}{z^2}- \frac{\delta_{ i j}}{z} (J_{SB}+J_F)\,, \\
M^{\pm}_i (z) F^{\pm}_j (0) &\sim \delta_{i j} \frac{M^{\pm}_i }{z}  \\
M^{\pm}_i (z) F^{\mp}_j (0) &\sim \frac{ \pm  F^{\pm} }{z}\,, \\
F^{\pm}_i (z) F^{\pm}_j (0) &\sim 0\,,   \\
F^{+}_i (z) F^{-}_j (0) &\sim \frac{\delta_{i j}}{z^2}+\frac{\delta_{i j} }{z}(J_F + J_{SB})\,.
\end{align}
}}

\rem{
To derive the last term in \eqref{eq:BOalg}, we used the relation 
\begin{align}
&: X_1X_2 Y_1 Y_2: +: X_1X_3 Y_1 Y_3: +: X_2 X_3 Y_2 Y_3: +\sum_{i=1}^3 :(\partial X_i) Y_i:
  \nonumber \\
& \qquad =3 : J_{SB}^2: -\frac{3}{2} \partial J_{SB} -T - \frac{1}{3} Q_B \left( b_1 (J^{(1)}_{\mathrm{sb}}-J^{(1)}_{f}-J^{(2)}_{f} ) +b_2 (J^{(1)}_{\mathrm{sb}}, + J^{(2)}_{sb} -J^{(2)}_{f} ) \right)
\end{align}
\begin{align}
J_{sbf,i}:=X_i Y_i+\psi_i \chi_i
\end{align}
\begin{align}
-3(J_{SB}+J_F)=J_{sbf,1}+J_{sbf,2}+J_{sbf,3}=3 J_{sbf,i}
\end{align}
\begin{align}
J_{sbf,1}&=J_{sbf,2}+ Q_{\rm BRST} \cdot {\sf b}_1 \\
J_{sbf,2}&=J_{sbf,3}+ Q_{\rm BRST} \cdot {\sf b}_2 \\
J_{sbf,1}&=J_{sbf,3}+ Q_{\rm BRST} \cdot ({\sf b}_1+{\sf b}_2) 
\end{align}
}

\rem{
\begin{align}
Z_{S^1 \times D^2}^{ (H),\widetilde{\mathcal{T}}^N_{\rm SQED}}&=Z_{S^1 \times D^2}^{(C), \mathcal{T}^N_{\rm SQED}}
=1+\left[1+a^2 -N\left(s+\frac{1}{s} \right) \right]q \nonumber
\\
 & 
+
\left\{
\begin{array}{l}
\qquad \sum_{n=2}^{N-1} c_n (s) +\sum  \,  q^n, \quad (N={\rm even}),\\
 \left( a+ \frac{1}{a}+\sum_{l=-N \atop l \neq 0}^{N} (-1)^l  a^{{\rm sign}(l)} \binom{N}{l}  s^l \right) q^{\frac{N}{2}}  +O(q^{\frac{N+1}{2}}), \, \, \, (N={\rm odd}).
\end{array}
\right. 
\end{align}
}



\bibliography{refs}

\end{document}